\documentclass[nofootinbib,twocolumn,prl]{revtex4}
\usepackage{graphicx}
\usepackage{amsfonts}
\usepackage{amssymb}
\usepackage{amsmath}
\usepackage{bm}
\begin{document}


\title{Detecting extra dimensions with gravity wave spectroscopy:\\
the black string brane-world }

\author{Sanjeev S.~Seahra, Chris Clarkson, Roy Maartens}

\affiliation{Institute of Cosmology \& Gravitation, University of
Portsmouth, Portsmouth~PO1~2EG, UK}

\setlength\arraycolsep{2pt}

\newcommand*{\Y}{{Y}_{lm}}
\newcommand*{\s}{ {}^{m}_{\,\,l} {\mathbb{S}} }
\newcommand*{\Vone}{ {}^{m}_{\,\,l} {\mathbb{V}}^{(1)}_a }
\newcommand*{\Vtwo}{ {}^{m}_{\,\,l} {\mathbb{V}}^{(2)}_a }
\newcommand*{\Tone}{ {}^{m}_{\,\,l} {\mathbb{T}}^{(1)}_{ab} }
\newcommand*{\Ttwo}{ {}^{m}_{\,\,l} {\mathbb{T}}^{(2)}_{ab} }
\newcommand*{\Tthree}{ {}^{m}_{\,\,l} {\mathbb{T}}^{(3)}_{ab} }

\newcommand*{\SSS}{  {\mathbb{S}} }
\newcommand*{\VVone}{  {\mathbb{V}}^{(1)}_a }
\newcommand*{\VVtwo}{  {\mathbb{V}}^{(2)}_a }
\newcommand*{\TTone}{  {\mathbb{T}}^{(1)}_{ab} }
\newcommand*{\TTtwo}{  {\mathbb{T}}^{(2)}_{ab} }
\newcommand*{\TTthree}{  {\mathbb{T}}^{(3)}_{ab} }

\newcommand*{\polar}[1]{ {}^{m}_{\,\,l} {\mathbb{P}}^{(#1)}_{AB} }
\newcommand*{\axial}[1]{ {}^{m}_{\,\,l} {\mathbb{A}}^{(#1)}_{AB} }

\newcommand*{\axpot}[2]{ V^\text{\tiny{#1}}_{#2} }

\newcommand*{\PP}[1]{ {\mathcal P}^{#1} }
\newcommand*{\Axial}[1]{ {\mathcal A}^{#1} }

\newcommand*{\polarlm}[1]{ {\mathcal P}_{lm}^{#1} }
\newcommand*{\scalarlm}[1]{ {\mathcal S}_{lm}^{#1} }
\newcommand*{\vectorlm}[1]{ {\mathcal V}_{lm}^{#1} }
\newcommand*{\tensorlm}[1]{ {\mathcal T}_{lm}^{#1} }

\newcommand*{\scalar}[1]{ {\mathcal S}^{#1} }
\newcommand*{\Vector}[1]{ {\mathcal V}^{#1} }
\newcommand*{\Tensor}[1]{ {\mathcal T}^{#1} }

\newcommand*{\di}{\partial}


\date{\today}

\begin{abstract}

Using the black string between two branes as a model of a
brane-world black hole, we compute the gravity wave perturbations
and identify the features arising from the additional
polarizations of the graviton. The standard four-dimensional
gravitational wave signal acquires late-time oscillations due to
massive modes of the graviton. The Fourier transform of these
oscillations shows a series of spikes associated with the masses
of the Kaluza-Klein modes, providing in principle a spectroscopic
signature of extra dimensions.

\end{abstract}

\maketitle


Black holes are central to our understanding of gravity, and are
expected to be key sources of gravity waves that should be
detected by the current and upcoming generation of experiments.
Such a detection will not only confirm the indirect evidence from
binary pulsars for gravity waves, but will also also allow us to
probe the properties of black holes and of gravity. In particular,
this will open up a new window for testing modifications to
general relativity, such as those arising from quantum gravity
theories. String theory for example predicts that spacetime has
extra spatial dimensions, so that the gravitational field
propagates in higher dimensions and has extra polarizations.
Recent developments in string theory indicate that Standard Model
fields may be confined to a four-dimensional `brane', while
gravity propagates in the full `bulk' spacetime. This has spurred
the development of brane-world models, such as Randall-Sundrum
(RS) type models, which can be used to explore astrophysical
predictions~\cite{rs}. RS type models have a five-dimensional bulk
with negative cosmological constant, so that the metric is warped
along the extra dimension. As a result, these models provide a new
approach to the hierarchy problem, dimensional reduction and
holography.

\begin{figure}[!t]
\begin{center}
\includegraphics[width=1\columnwidth]{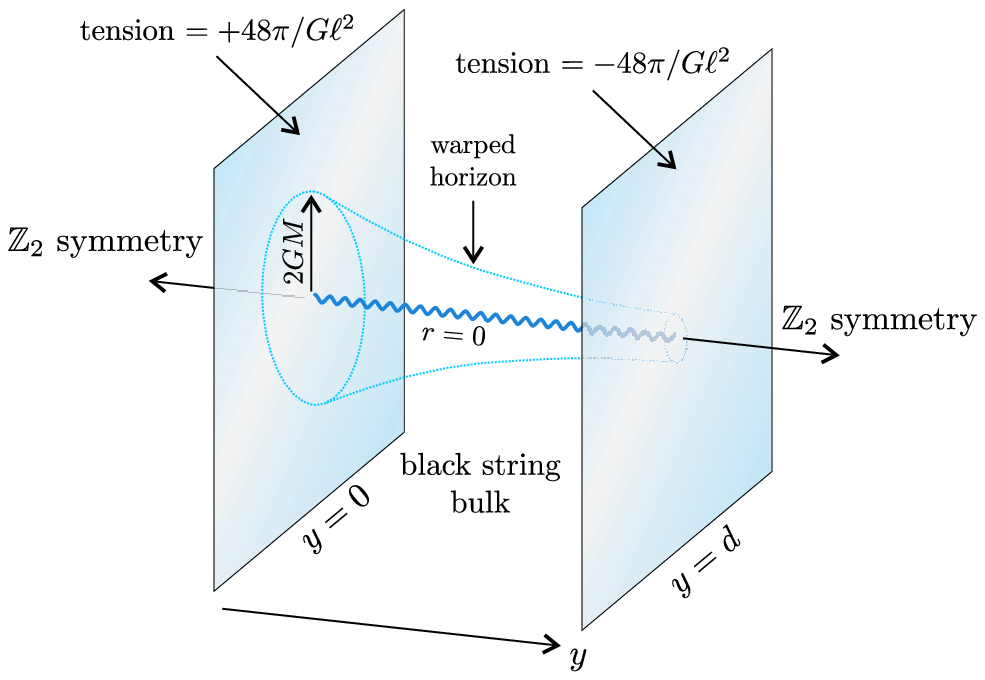}\\
\caption{Schematic of the black string}\label{F1}
\end{center}
\vspace{-7mm}
\end{figure}

The nature of black holes that form by gravitational collapse on
an RS brane is only partly understood~\cite{rs,ktn}, and no exact
solution is known for a black hole localized on one brane. If
there is a second `shadow' brane, the black string may be used to
model large black holes on the visible brane, when the horizon on
the brane is much greater than the extent of the horizon into the
bulk~\cite{ks}. The black string reproduces the Schwarzschild
metric on the visible brane but is not confined to the brane,
since there is a line singularity at $r=0$ into the extra
dimension (see Fig.~\ref{F1}). The shadow brane can also introduce
an infra-red cut-off to shut down the Gregory-Laflamme (GL)
instability of the black string at long wavelengths~\cite{gl}. If
the shadow brane is close enough to the visible brane for a given
black hole mass $M$, or if $M$ is large enough for a given brane
separation $d$, then $GM{\rm e}^{-d/\ell}/\ell$ is above a
positive critical value and the GL instability is removed (see
below). This is the background model that we perturb.

The brane separation is constrained from above by stability
requirements. It is also constrained from below. This follows
since the brane separation is a massless degree of freedom, felt
on the visible brane as a `radion' field, so that the low-energy
effective theory on the visible brane is of Brans-Dicke type,
with~\cite{gt} $\omega_{\rm bd}=3( {\rm e}^{2d/\ell}-1)/2$, where
$\ell$ is the bulk curvature radius. The shadow brane must be far
enough away that its gravitational influence on the visible brane
is within observational limits. Solar system observations impose
the lower limit~\cite{wy} $\omega_{\rm bd} \gtrsim 4 \times 10^4$,
so that $d/\ell\gtrsim 5$. The allowed region in parameter space
is shown in Fig.~2. Table-top tests of Newton's law impose the
constraint $\ell\lesssim 0.1\,$mm. This upper limit defines a mass
$0.1\,\mbox{mm}/2G \sim 10^{-7}M_\odot$, so that for astrophysical
black holes it follows that $2G M\gg \ell$ is easily satisfied:
$GM/\ell \gtrsim 10^7(M/M_\odot)$.

\begin{figure}[!t]
\begin{center}
\includegraphics[width=0.9\columnwidth]{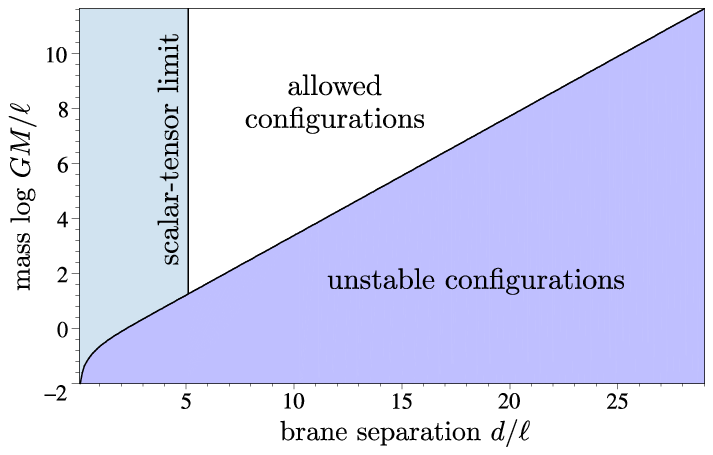}\\
\caption{The allowed region in parameter space.  The stability
boundary is well approximated by $GM/\ell = 0.1122 {\rm
e}^{d/\ell}$.}\label{F2}
\end{center}
\vspace{-7mm}
\end{figure}

The 5D black string is a solution of the Einstein equations
$G_{AB} = 6 \ell^{-2} g_{AB},$ with metric
\begin{eqnarray} \label{black string metric}
ds^2 & = & a^2 \left( -f\,dt^2+f^{-1}\,dr^2+r^2\,d\Omega^2 \right)
+ dy^2,
\end{eqnarray}
where $f(r)  =   1 -2GM/r, \quad a(y) = {\rm e}^{-|y|/\ell}$, and
the branes are at $y=0,\,d$.
Metric perturbations satisfy
\begin{equation}\label{general perturbation equation}
\Box h_{AB} +\nabla_A \nabla_B h^C{}_C - 2 \nabla^C \nabla_{(A}
h_{B)C} - 8 \ell^{-2} h_{AB} = 0,
\end{equation}
and gauge choices may be made to reduce the degrees of freedom to
5, the number of polarizations of the 5D graviton.

As in standard black hole perturbation theory \cite{nol}, we use
spherical harmonics $Y_{lm}$ and their gradients to construct a
basis $\mathbb{T}^{(j,lm)}_{AB}$ for $h_{AB}$, with $j = 1 \ldots
15$. The metric perturbation is naturally split into two decoupled
parts: polar [parity $(-1)^l$] and axial [parity $(-1)^{l+1}$].
Suppressing the $(lm)$ indices, the expansion coefficients are of
the form $\mathcal{C}^{(j)} = {\rm e}^{i \omega
\tau}H^{(j)}(r,y)$, where $\tau = t/GM$ and $\omega$ is a
dimensionless frequency. We use two gauges, which are related by
quadrature. We generalize the 4D Regge-Wheeler (RW) gauge by
setting the coefficients of the most complicated harmonic tensors
equal to zero, which makes the remaining coefficients gauge
invariant.  The RS gauge~\cite{gt}, given by $h^A{}_{A} = 0 =
\nabla^B h_{AB} = h_{Ay}$, has the advantage that the expansion
coefficients are separable: $\mathcal{C}^{(j)} = {\rm e}^{i \omega
\tau} W^{(j)}(r)Z(y)$, where \cite{gl}
\begin{equation}\label{y profile}
    a^2 Z'' -2(aa')'Z = -m^2 Z .
\end{equation}
Here, $m$ is the effective mass on the visible brane of the
Kaluza-Klein (KK) mode of the 5D graviton. The bulk wave functions
$Z(y)$ are the same as those for Minkowski branes (even though the
bulk is no longer anti de Sitter), and the solution for $m > 0$ is
$Z(y)=B_2(m\ell/a)$, where $B_2$ is a linear combination of Bessel
functions.

Neglecting brane bending, the boundary conditions at $y=0\,,d$ in
RS gauge are $\di_y(a^{-2} h_{\alpha\beta} )=0$. When $m=0$, the
solution is $Z \propto a^2$. For this zero-mode, the metric
perturbations reduce to those of a 4D Schwarzschild metric, as
expected.
For $m \ne 0$, the boundary conditions lead to a discrete tower of
KK mass eigenvalues,
\begin{equation}
m_n=(z_n/\ell){\rm e}^{-d/\ell}\,,
\end{equation}
where $Y_1(m_n\ell)J_1(z_n)=J_1(m_n\ell)Y_1(z_n)$~\cite{rs}. The
GL instability exists in the range $0<m<m_{\rm crit}\approx
0.4/GM$~\cite{gl}. If $m_1>m_{\rm crit}$, then the instability is
avoided, and $m_1=m_{\rm crit}$ defines the stability curve in
Fig.~\ref{F2}.

\paragraph*{Radial master equations.}

We generalize the standard 4D analysis to find radial master
equations for a reduced set of variables, for all classes of
perturbations. In the 5D case, the KK modes are governed by
coupled master equations.
These can be written as matrix-valued Schr\"odinger-like
equations,
\begin{equation}\label{schro}
 -\frac{\mathrm d^2}{\mathrm d x^2} \bm\Psi
+\mathbf V\bm\Psi = \omega^2\bm\Psi,
\end{equation}
where $\mathbf{V}$ is the potential matrix,
$x=\rho+2\ln{(\rho/2-1)}$, and $\rho = r/GM$. For spherical
(s-wave) polar perturbations, there is only one master variable,
and
\begin{equation}
\!\frac{V_\mathrm{s}}{f}\!=\!{\frac{
{\mu}^{6}{\rho}^{9}+6\,{\mu}^{4}{
\rho}^{7}-18\,{\mu}^{4}{\rho}^{6}-24\,{\mu}^{2}
{\rho}^{4}+36\,{\mu}^{2}{\rho }^{3}+8 }{{\rho}^{3} \left(
{\mu}^{2}{\rho}^{3}+2 \right)^{2}}}\!,
\end{equation}
where $\mu=GMm$. For $l \ge 2$ axial perturbations, there are two
master variables and the potential matrix in the RW gauge is
\begin{equation}
\frac{\mathbf{V}_\mathrm{a}}{f} =\mu^2\bm I+ {\rho^{-3}} \left[
\begin{array}{cc}
     {l(l+1)}{\rho} - 6
    & {\mu^2} \\ 4\rho^3 &   {l(l+1)}{\rho}
\end{array}
\right].
\end{equation}
The diagonal elements of $\mathbf{V}_\mathrm{a}$ are identical to
the RW potentials for spin-2 and spin-1 excitations of 4D
Schwarzschild, plus a mass term. Massive KK modes of the 5D
graviton include spin-1 perturbations.  (For $\mu = 0$, the spin-1
contribution is pure gauge.) For the other types of perturbations,
we find that the $l=1$ and $l \ge 2$ polar perturbations are
governed by two and three master variables respectively, while
there is only one master variable for the axial p-wave.

\paragraph*{Stability.}

\begin{figure}[!b]
\begin{center}
\includegraphics[width=\columnwidth]{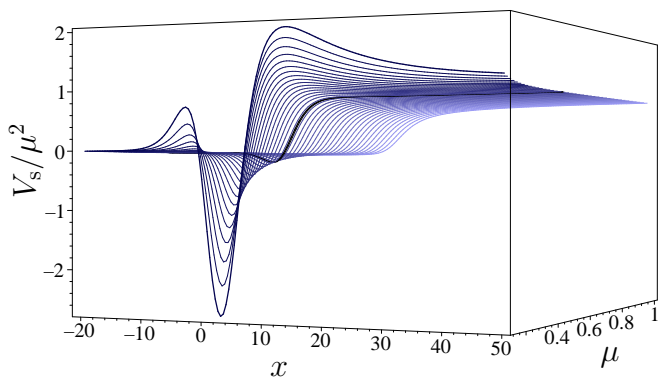}\\
\caption{The s-wave potential for varying $\mu$. The critical
potential is indicated by the heavy line.  For $\mu <
\mu_\textrm{crit}$, there is a bound state with $\omega^2 < 0$,
indicating an instability.}\label{F3}
\end{center}
\vspace{-7mm}
\end{figure}

All master equations can be cast in the form
${\mathcal{L}}(\mu,\omega) \mathbf{\Psi} = 0$, where ${\mathcal
L}$ is a linear differential operator. Stability holds if
${\mathcal L}$ is positive definite when the boundary conditions
$\mathbf{\Psi} =0$ are imposed at $x=\pm\infty$, and $\omega^2 <
0$~\cite{nol}. We have verified the positivity of ${\mathcal L}$
analytically for all classes of axial perturbation.
For the $l \ge 1$ polar case
we performed a numerical search for solutions satisfying the above
boundary conditions in the relevant region of $(\mu,\omega)$
parameter space. No solutions were found, suggesting that no
instability exists.
The opposite is true for the s-wave case. The reason lies in the
potential, which is shown in Fig.~\ref{F3} for different values of
$\mu$. For small $\mu$, there is a potential well that can support
a normalizable bound state with $\omega^2 < 0$, which implies that
${\mathcal L}$ is not positive definite. This well vanishes for
large $\mu$, suggesting that an instability exists for all modes
with $0<\mu < \mu_\text{crit}$. We used a standard WKB-inspired
phase integral analysis~\cite{and} to develop a new improved
derivation of the GL instability. Our approach leads to an
accurate determination, $\mu_\textrm{crit} = 0.4301$, in agreement
with recent results from full numerical relativity~\cite{cho}.
Stability is achieved if the mass of the first KK mode is such
that $GMm_1 > \mu_\textrm{crit}$ (see Fig.~2).

\paragraph*{Gravity wave signals.}

\begin{figure}
\begin{center}
\includegraphics[width=\columnwidth]{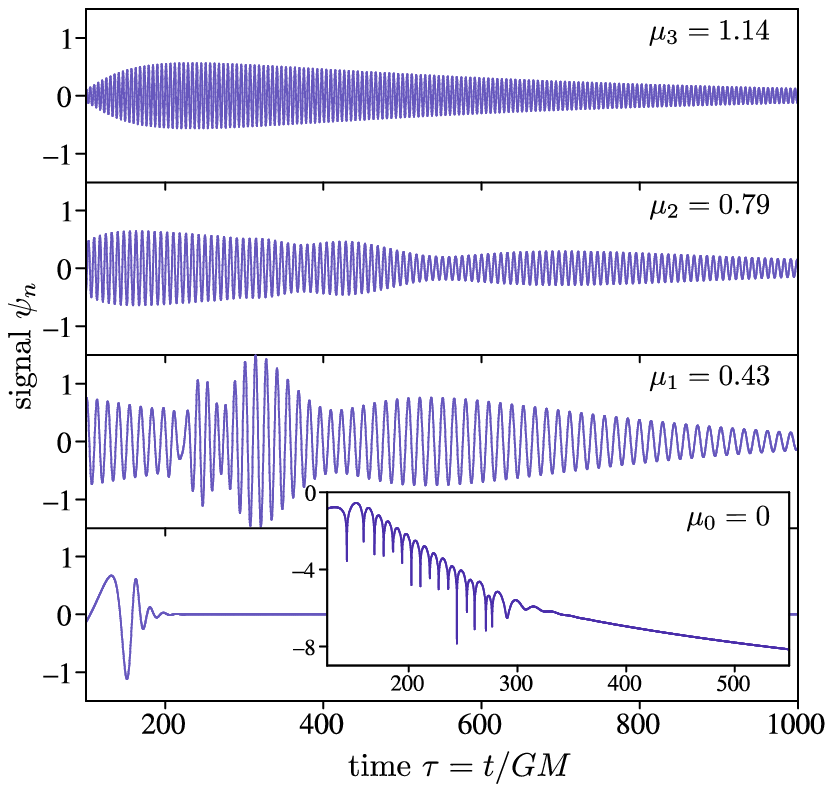}\\
\caption{Gravity wave signals associated with 3 massive KK modes
and the 0-mode of $l=2$ axial perturbations, for a marginally
stable black string.  The inset shows the zero-mode signal on a
logarithmic scale.}\label{F4}
\end{center}
\vspace{-7mm}
\end{figure}

Perturbations of a stable black string describe gravitational
radiation that may in principle be observable by current or future
detectors. The total gravity wave signal at the observer
($x=x_{\rm obs}$) is a superposition of the waveforms
$\psi_n(\tau)$ associated with the mass eigenvalues $m_n$ of
Eq.~(\ref{y profile}). In Fig.~\ref{F4}, we present signals
associated with the four lowest masses for a marginally stable
black string, for $l=2$ axial modes. These are obtained by
re-introducing time dependence into the Schr\"odinger
equation~(\ref{schro}), via $\omega \rightarrow -i \, \di/\di
\tau$, and numerically integrating the resulting PDE with
$\mathbf{V} = \mathbf{V}_\mathrm{a}$. We assume static Gaussian
initial data centered about $x=50$ and an observer at
$x_\mathrm{obs} = 100$. Because the master equations reduce to the
4D RW equation when $\mu = 0$, the bottom curve exhibits the
damped single-frequency (quasi-normal) ringing followed by a
power-law tail familiar from the 4D analysis \cite{nol}.  The
massive mode signals are quite different~\cite{bkrk}, showing much
less damping than the massless mode, and late-time monochromatic
oscillations instead of a featureless power-law tail. This is
reminiscent of the behaviour of massive scalar fields in 4D
spherically symmetric spacetimes, where a detailed Green's
function analysis shows~\cite{Koy02} that the late-time signal is
$\propto t^{-5/6} \sin \, mt$, irrespective of $l$.

\begin{figure}[!t]
\begin{center}
\vspace*{2mm}
\includegraphics[width=\columnwidth]{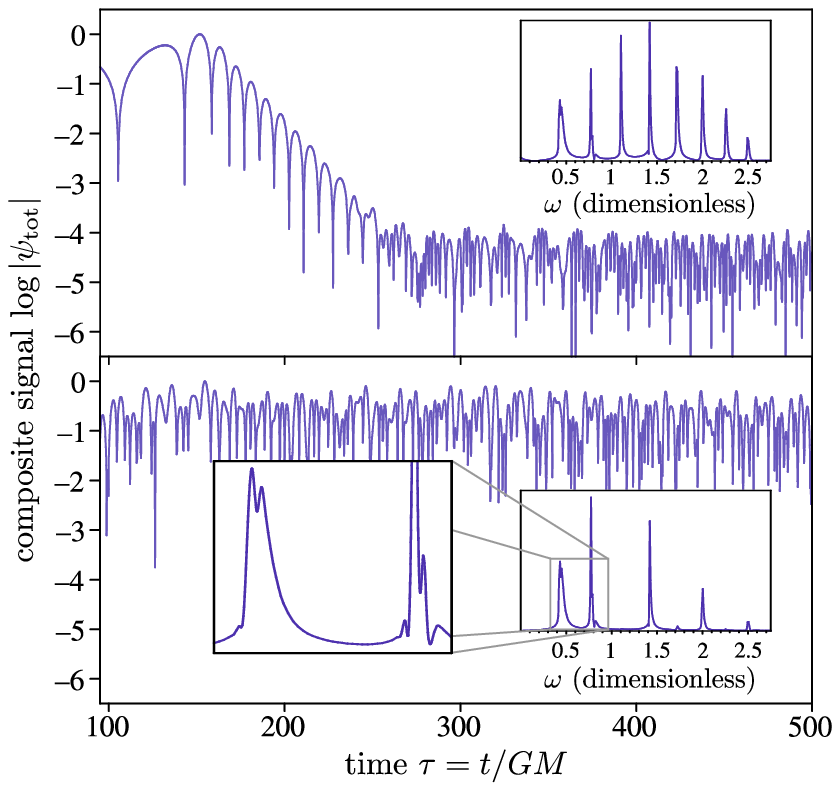}
\caption{Composite $l=2$ axial gravity wave signal (normalized)
from the 9 lowest mass modes of a marginally stable black string.
The upper panel corresponds to `truncated zero-mode' bulk initial
data with $d/\ell = 6$, the lower panel to Gaussian bulk initial
data with $d/\ell = 20$. Insets show the Fourier transform of the
respective waveforms for $300\leq \tau \leq 1000$.  The percentage
of energy in the 4D zero-mode part of the signal is 99.96\,\%
(upper) and $2.281 \times 10^{-13}$\,\% (lower).}\label{F5}
\end{center}
\vspace{-9mm}
\end{figure}

To reconstruct the total waveform $\psi_\mathrm{tot}$ for the
$l=2$ axial case, as measured by an observer on the visible brane,
we need to specify an initial profile $F$ for the perturbation in
the bulk. Using a conformal bulk coordinate $\xi = a^{-1}$, and
assuming that $F$ and $Z_n$ are appropriately normalized, we have
\begin{equation}
\psi_\mathrm{tot}(\tau) = \sum_n \, c_n Z_n(0) \psi_n(\tau),
  \,\, c_n = \int_{1}^{{\rm e}^{d/\ell}} \!\xi F Z_n d\xi,
\end{equation}
where $\sum_n\, c_n^2=1$. Hence, $c_n^2$ is the fractional energy
in the $n$-mode. Massive KK modes introduce striking new features
in the gravity wave signal that are potentially observable, as
shown in Fig.~\ref{F5}. The strength of the new features is
sensitive to the initial data $F(\xi)$. Here we simply illustrate
the possibilities via two forms of $F(\xi)$ that correspond to
very different situations.

In Fig.~\ref{F5}, the upper signal has $d/\ell = 6$~\cite{note},
and the initial data $F(\xi)$ corresponds to the zero-mode
$1/\xi^2$ with a cut-off applied one eighth of the distance to the
shadow brane. This data is motivated by the results of numerical
calculations of the bulk gravitational field around a small
brane-localized black hole~\cite{ktn}. If the black hole mass is
small enough to be in the unstable region of Fig.~2, we may think
of the black hole as the endpoint of a GL
instability~\cite{ktn,cho}. The initial data is then thought of as
arising from an encounter between this black hole and the much
heavier black string.
We find that in this case, the coefficients of the massive mode
signals are of order $10^{-4}$, which suggests that the total
waveform will only exhibit minor deviations from the GR
prediction. This is certainly true for early times, where the
signal is dominated by ordinary 4D Schwarzschild quasi-normal
ringing (\emph{cf.}~the inset of Fig.~\ref{F4}). However, as can
be inferred from the individual massive waveforms in Fig.~4, the
late-time signal is very different from the 4D case--it is both
oscillatory and very lightly damped. The inset shows the Fourier
transform of the signal for late times. There are eight discrete
peaks in the spectrum, corresponding to each of the eight non-zero
mass modes.

In the lower panel, we take $d/\ell = 20$ and Gaussian initial
data in the bulk centered halfway between the branes (if the
center is moved closer to the shadow brane, the results are
qualitatively similar). This could correspond to an event which
`mainly takes place in the bulk', such as the merger of two black
strings. In this case, the zero mode waveform is dominated by the
massive mode signals, and the waveform is very different from the
4D case. The contributions from the odd modes $m_3$, $m_5$ and
$m_7$ are suppressed, which results in a late-time transform with
five principal peaks in the inset. These peaks in fact display a
degree of fine structure, as shown in the blow-up, that
deserves
further investigation.

\paragraph*{Conclusions.}

Will the consideration of other types of perturbations or
different multipoles in some way contaminate the spectroscopic
signal? We find that in all cases, as $r \rightarrow \infty$,
master variables with $\mu\ne0$ behave as massive fields
propagating on Schwarzschild spacetime, which is enough to
guarantee the slowly-decaying oscillating tail. Massless modes for
$l > 2$ are known to be subject to more damping than for $l = 2$,
so if the massive signal dominates the latter for late times, it
will also dominate the former.

Can the massive signal be resolved by realistic gravity wave
detectors? There are two separate issues: the frequency of the
massive modes and their relative amplitude. From the formulas
above, one finds that the discrete frequencies in the late-time
tail are
\begin{equation}
f_n = z_n {\rm e}^{26.9-d/\ell}
(0.1\,\text{mm}/\ell)\,\,\text{Hz.}
\end{equation}
For $d/\ell \gtrsim 5$, $z_n$ is well approximated by
$J_1(z_n)=0$. Note that unlike 4D black hole quasinormal modes,
these frequencies are {\em independent} of the mass $M$. Taking
$\ell = 0.1\,$mm and $d /\ell=24$, results in $f_1 \sim 100\,$Hz,
which is ideal for a LIGO detection of extra dimensions. For $d
/\ell= 33$, we find an optimal LISA signal, with $f_1 \sim
0.01\,$Hz. For these parameters, the lower limits on black string
masses are $M>100M_\odot$ and $M>10^6M_\odot$, respectively.

In regard to the amplitudes, the most optimistic type of events
involve bulk-based initial data, i.e., black string mergers. Since
the zero-mode energy is small in these situations, we expect the
strength of the massive mode oscillations to be comparable to the
quasinormal ringing amplitude in the analogous 4D case. The
situation is more delicate for brane based initial data, where the
zero mode-tends to dominate the signal. We find that the relative
massive amplitude tends to increase with decreasing $d/\ell$.
However, small $d/\ell$ implies large $f_n$, so the best prospect
of seeing this type of signal lies with high frequency detectors.


The discrete nature of the late-time Fourier transform of the
black string waveform is the most important observable feature of
this model. Its detection would provide clear evidence of extra
dimensions. It would also give the direct spectroscopic
measurement of the KK masses $m_n$, which in turn provides
information about $d$ and $\ell$. Although we have analyzed a
specific model of brane-world black holes, we expect that
qualitatively similar features will arise for other models with
compact extra dimensions, since they all have a discrete tower of
massive KK modes. Furthermore, since the massive modes travel
below light-speed, there will be potentially observable time-delay
in their arrival, which could be of order seconds or longer for
distant sources. Also, the light damping of KK modes means that
they may have a significant integrated contribution to the
stochastic gravity wave background.

\paragraph*{Acknowledgements} SSS is supported by NSERC, CC and RM by
PPARC. We thank C Kiefer, B Kol, K Koyama, D Langlois and J Soda
for discussions.


\begin{thebibliography}{99}
\bibitem{note}
For numerical convenience we choose the marginally stable case, so
that $M\ll M_\odot$ for $d/\ell=6$, but the results will be
qualitatively similar for $M \gtrsim M_\odot$.
\end{thebibliography}
\end{document}